\RequirePackage{fix-cm}
\documentclass[twocolumn]{svjour3}          
\smartqed  
\usepackage{graphicx}
\usepackage{mathptmx}      
\usepackage{amssymb}
\usepackage{amsmath}
\usepackage{mathrsfs}
\usepackage{subfig}
\usepackage{tikz}
\usepackage{pgfplots}

\usepackage{epstopdf}

\usepackage{natbib}
\usepackage{pstricks}
\usepackage[breaklinks]{hyperref}
\usepackage{doi}
\usepackage{multirow}

\newcommand{\ie}{\emph{i.e.}~}
\newcommand{\mb}\boldsymbol
\newcommand{\pdiff}[2]{\frac{\partial #1}{\partial #2}}
\newcommand{\rdiff}[2]{\frac{\mathrm{d} #1}{\mathrm{d} #2}}
\newcommand{\mdiff}[2]{\frac{\mathrm{D} #1}{\mathrm{D} #2}}

\journalname{Biomechanics and Modeling in Mechanobiology}

\begin{document}

\title{A model for one-dimensional morphoelasticity and its application \\
to fibroblast-populated collagen lattices
}

\titlerunning{A $1$-D morphoelastic model for fibroblast-populated collagen lattices}  

\author{Shakti N.~Menon \and
        Cameron L.~Hall \and
        Scott W.~McCue \and
        D.~L.~Sean McElwain}


\institute{Shakti N.~Menon \at
              \emph{Present address:} The Institute of Mathematical Sciences, CIT Campus, Taramani,
              Chennai 600113, India\\
              School of Mathematical Sciences, Queensland University of Technology,
              Brisbane QLD 4001, Australia\\
              Tissue Repair and Regeneration Program, Institute of Health and Biomedical Innovation,
              Queensland University of Technology, Brisbane QLD 4001, Australia
            \and
          Cameron L.~Hall \at
          	  \emph{Present address:} Mathematics Applications Consortium with Science and Industry, University of Limerick, Castletroy, Limerick, V94 T9PX, Ireland \\
              Oxford Centre for Industrial and Applied Mathematics, Mathematical Institute,
              University of Oxford, 24-29 St Giles', Oxford OX1 3LB, UK
          \and
          Scott W.~McCue \at
              School of Mathematical Sciences, Queensland University of Technology,
              Brisbane QLD 4001, Australia\\
              \email{scott.mccue@qut.edu.au}
          \and
          D.~L.~Sean McElwain \at
              School of Mathematical Sciences, Queensland University of Technology,
              Brisbane QLD 4001, Australia\\
              Institute of Health and Biomedical Innovation,
              Queensland University of Technology, Brisbane QLD 4001, Australia
}

\date{Accepted: 30 April 2017}

\maketitle


\begin{abstract}
\textcolor{black}{The mechanical behaviour of solid biological tissues has long been described using models based on classical continuum mechanics. However, the classical continuum theories of elasticity and viscoelasticity cannot easily capture the continual remodelling and associated structural changes of biological tissues. Furthermore, models drawn from plasticity theory are difficult to apply and interpret in this context, where there is no equivalent of a yield stress or flow rule. In this work, we describe a novel one-dimensional mathematical model of tissue remodelling based on the multiplicative decomposition of the deformation gradient. We express the mechanical effects of remodelling as an evolution equation for the ‘effective strain’, a measure of the difference between the current state and a hypothetical mechanically-relaxed state of the tissue. This morphoelastic model combines the simplicity and interpretability of classical viscoelastic models with the versatility of plasticity theory. A novel feature of our model is that while most models describe growth as a continuous quantity, here we begin with discrete cells and develop a continuum representation of lattice remodelling based on an appropriate limit of the behaviour of discrete cells. To demonstrate the utility of our approach, we use this framework to capture qualitative aspects of the continual remodelling observed in fibroblast-populated collagen lattices, in particular its contraction and its subsequent sudden re-expansion when remodelling is interrupted.}
\keywords{morphoelasticity \and biomechanics \and tissue plasticity \and fibroblast-populated collagen lattices}
\subclass{74L15 \and 92C10 \and 74D10}
\end{abstract}


\section{Introduction}
\label{S:Intro}

In this study, we present a one-dimensional mathematical model of biological tissue remodelling, based on the multiplicative decomposition of the deformation gradient.  An important feature of our model is that the mechanical effects of remodelling are expressed in terms of an evolution equation for the `effective strain' -- a measure of the difference between the current state and a hypothetical mechanically-relaxed state of the tissue.  This morphoelastic model combines the simplicity and interpretability of classical viscoelastic models with the versatility of plasticity theory. To demonstrate its utility, we show that this model can quantitatively capture aspects of the mechanical behaviour of fibroblast-populated collagen lattices reported in previous experiments.

As biological tissues deform continuously when subjected to mechanical forces, their physical behaviour is often modelled using classical continuum mechanics \citep{Murray2001}. However, unlike classical solids, living tissues may contain cells that can modify the fundamental mechanical properties of their physical environment. There are a number of processes, most notably tissue growth, in which cells cooperatively alter the tissue structure, changing the relationship between stress and deformation \citep{Chen2000}. Indeed, many tissues undergo a continual process of internal revision and mechanical restructuring, often referred to as `remodelling' \citep{Taber1995}, in which physical properties of the material, including anisotropy and stiffness, evolve over time. This active remodelling is thought to be significant in a wide range of biological processes such as embryo development and morphogenesis (see \citet{Patwari2008} for a recent discussion). In particular, it is well known that fibroblast cells, which are found in the stroma of numerous tissues, actively remodel the surrounding extracellular matrix (ECM) by synthesising and reorganising collagen fibres, and that this remodelling is essential for tissue homeostasis and for wound repair \citep{Grinnell2003,MajnoCells}.  Remodelling also affects the mechanical stresses experienced by cells in a tissue, which can subsequently modify aspects of cell behaviour. For instance, fibroblasts are known to change their morphology \citep{Gabbiani2003,Tamariz2002,Tomasek2002} and phenotype \citep{Amadeu2003,Gabbiani2003,Tomasek2002} in response to external mechanical cues. Most significantly, the mechanical stresses experienced by fibroblasts during the wound healing process stimulate them to differentiate into more contractile forms: protomyofibroblasts and myofibroblasts \citep{Desmouliere2005,Gabbiani1972,Tomasek2002}, which play a major r\^{o}le in the subsequent contraction of the wound. The interplay between mechanical stress and active remodelling is hence critically important in wound healing, as excessive contraction is known to lead to pathologies, such as hypertrophic scars and contractures \citep{Roseborough2004,Enoch2005,Murphy2011b}.

\begin{figure}[t]
\begin{center}
\includegraphics[scale=1.0]{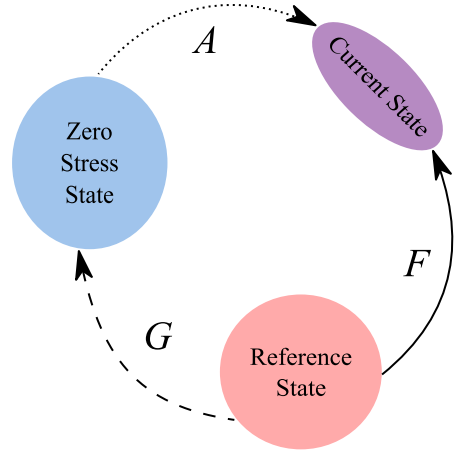}
\caption{The relationship between the reference state, the current state and the zero stress state. In our notation, $\mb{F}$ represents the overall deformation gradient from the reference state to the current state, while the plastic/growth component $\mb{G}$ represents the deformation from the reference state to the zero stress state and the elastic component $\mb{A}$ represents the deformation from the zero stress state to the current state.
}\label{F:ZSS}
\end{center}
\end{figure}

Since the first descriptions of the kinematics of biological growth the 70s and early 80s \citep{Taber1995,Humphrey2003}, several theoretical frameworks have been proposed to capture the dynamics of remodelling. A very significant advance in this direction was made in the mid 1990s, when \citet{Rodriguez1994} and \citet{CookThesis} independently developed a mathematical framework for remodelling that utilizes the \textit{multiplicative decomposition of the deformation gradient} -- an idea that dates back to the 50's~\citep{Bilby1957,Kroner1958,Kroner1959}, and which was formalized in the 60's by \citet{Stojanovic1964,Stojanovic1970} and \citet{Lee1969}. In this paper, we use the notation developed by Goriely and coworkers \citep{Goriely2007,Goriely2008,Vandiver2009,Goriely2011}, where $\mb{A}$ and $\mb{G}$ represent the elastic and plastic/growth parts of the deformation gradient, respectively. The decomposition of the deformation gradient tensor thus yields the expression
\begin{equation}
\mb{F} = \mb{A} \, \mb{G} \,.
\label{MultDecomp}
\end{equation}
The physical interpretation of this decomposition is depicted in Figure \ref{F:ZSS}: $\mb{G}$ is the deformation gradient tensor associated with a hypothetical deformation from the fixed reference state to a state where all internal stresses are relieved, while $\mb{A}$ represents the deformation from this `zero stress state' to the current state. For a purely elastic material, the fixed reference state will also be the zero stress state, and we find that $\mb{G} \equiv \mb{I}$. However, plastic flow or remodelling enables the zero stress state, and hence $\mb{G}$, to evolve. A detailed discussion of the fundamental concepts that underlie the multiplicative decomposition of the deformation gradient, is presented in Appendix~\ref{S:Theory}.

This approach not only provides a clear and coherent way of understanding growth, but also leads to a natural way of describing ``residual stress'' -- stress that persists even when all loads are removed. Although \citet{FungBiomechMPLT} had noted much earlier that some tissues (such as arteries) are structured so that it is impossible for the entire tissue to be free of residual stresses unless cuts are made, limited attempts had been made to describe these stresses mathematically. An alternative paradigm was suggested by \citet{Goriely2007}, who coined the term \emph{morphoelasticity} to describe the combination of elastic and ``plastic'' changes that are the result of biological growth and remodelling. Despite some similarities, morphoelasticity is quite distinct from classical plasticity. For example, morphoelastic remodelling will generally occur throughout a tissue, not just in those regions where a yield stress is exceeded. Moreover, morphoelasticity can involve changes to the total mass of a tissue (in tissue growth, for example) and/or increases in internal energy, while plastic flow is always mass-conserving and dissipative. This framework has been recently used to mathematically describe aspects of wound healing, namely wound contraction and scar formation~\citep{Yang2013} and dermal wound closure~\citep{Bowden2016}, as well as growth in other biological tissues such as axons~\citep{Garcia2016}.

\begin{figure}[t]
\center{\includegraphics[width=0.45\textwidth]{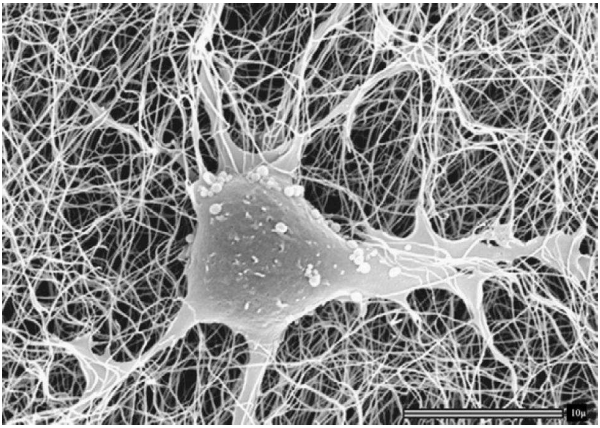}}
\caption{Scanning electron microscope image of a human fibroblast interacting with 3D collagen matrices. Here, the fibroblast exhibits a dendritic structure, indicating that the environment is unstressed. The scale in the figure represents $10$ microns. (Reprinted from \emph{Advanced Drug Delivery Reviews} 59(13), Rhee S. and Grinnell F., \textit{Fibroblast mechanics in 3D Collagen matrices}, pg 1299-1305, Copyright (2007) with permission from Elsevier).}
\label{fig:SEM}
\end{figure}
While remodelling is important in a wide range of phenomena \emph{in vivo}, surprisingly few \emph{in vitro} experiments have been developed to explicitly study the macroscopic consequences of remodelling. A notable counter-example is the investigation of the contraction of fibroblast-populated collagen lattices (FPCLs) -- cultured fibroblasts embedded in (or placed on top of) three-dimensional ($3$-D) collagen matrices. \textcolor{black}{It has long been known that FPCLs can contract to a small fraction of their initial size within a few days \citep{Bell1979}. This contraction was first observed to be `permanent'  by Grinnell and coworkers  \citep{Grinnell1984,Guidry1985,Guidry1986}, who validated this finding by performing several experiments with varying numbers of fibroblast cells that were placed on top of collagen lattices of different initial densities.} Most fibroblasts did not invade the lattice, and were instead found to spread over its surface while reorganising proximal collagen fibres in the direction of spreading. It was thus proposed that the reorganisation of the lattice away from the cells was chemically mediated by the secretion of cell-binding factors such as fibronectin and proteoglycans \citep{Grinnell1984}. It was also observed that the addition of cytochalasin D, which suppresses gel reorganisation by inhibiting cell motility, resulted in a partial re-expansion of the gel \citep{Guidry1985,Guidry1986}. The relative magnitude of the re-expansion was found to be smaller in gels that were contracted by fibroblasts for a greater period of time, which suggested that the collagen gels were first physically reorganised by fibroblasts and then stabilised by the continued presence of these cells.

In this work, we develop a mathematical description of the contraction of FPCLs that incorporates the mechanical effects of remodelling.  In Sec.~\ref{section2} we provide a detailed summary of FPCLs and the different mechanisms by which they can permanently contract. Then, in Sec.~\ref{S:application}, we construct an expression for the rate of morphoelastic contraction of an FPCL, based on a plausible microscopic mechanism of cells rearranging the fibres of the lattice. On varying the two free parameters of this simplified model, we quantitatively replicate features of previous experiments on contracting FPCLs. Finally, in Sec.~\ref{S:discussion} we discuss how our framework is significantly advantageous in comparison to many other approaches to describing remodelling in biological tissues, and detail possible extensions to our model.


\section{Fibroblast-populated collagen lattices}
\label{section2}

\subsection{History and classification}
\label{S:IntroFPCL}

FPCLs were developed by \citet{Elsdale1972}, who used them as a means of investigating fibroblast behaviour in a setting that closely resembles their natural environment. A scanning electron microscope image of a fibroblast embedded in a collagen lattice, taken from \citet{Rhee2007}, is shown in Fig.~\ref{fig:SEM}. Experiments on such lattices can provide insight into the mechanical interactions between fibroblasts and surrounding collagen fibrils. Since first being developed, these lattices have been used to study the traction forces exerted by fibroblasts in mechanically-relaxed~\citep{Bell1979,Bellows1981,Ehrlich1990} in mechanically-loaded environments~\citep{Guidry1985,Hinz2001,Mudera2000}. FPCLs have also been used to investigate the effects of various growth factors on fibroblasts \citep{Grinnell1999,Schreiber2001} as well as the behaviour of individual fibroblasts as they contract their environment while moving through the ECM \citep{Roy1997,Roy1999}. Further details on the range of experiments involving FPCLs, and their clinical utility, are provided in the reviews by \citet{Dallon2008}, \citet{Grinnell2003}, and \citet{Ehrlich2013}.

FPCLs are typically classified according to the mechanical set-up that is used to create them. By this reckoning, there are three main types of FPCLs: free-floating, attached and stress-relaxed.

\textit{Free-floating FPCLs} were introduced by \citet{Bell1979} and are prepared by polymerising a collagen gel with fibroblasts. This could be done either in a bacteriological dish, to which the gel adheres poorly, or in a tissue culture dish, in which case the gel has to be detached after a certain time. The image displayed in Fig.~\ref{fig:FPCL}, taken from \citet{Kelynack2009}, shows an example of a free-floating FPCL. In such lattices, the fibroblasts project a dendritic network of extensions and the tension is distributed isotropically \citep{Grinnell2003b,Rhee2007}.

Fibroblasts in free-floating lattices can generate significant traction forces \citep{MajnoCells,Grinnell2000}, which reorganise the matrix and lead to contraction. While this reorganisation does not orient collagen fibrils in any particular direction, it can still cause these lattices to contract to as little as a tenth of their initial lateral (or vertical) extent \citep{Bell1979,Steinberg1980,Grinnell1984,Guidry1985}, even in the absence of protomyofibroblast cells, which can exert greater forces. This contraction gives rise to a mechanically relaxed tissue that resembles dermis, and it has hence been proposed that such lattices can be used to describe the earliest stages of wound healing, before inflammation and tissue stress have activated the differentiation of fibroblasts into myofibroblasts \citep{Grinnell1994}. Such FPCLs are observed to remain disk-shaped throughout the contraction process, which involves a reduction in thickness as well as diameter, although the edges of the disk are observed to eventually curl up \citep{Bell1979}.

\textit{Attached FPCLs} are fibroblast-populated lattices that are polymerised in a tissue culture dish, to which the gel attaches firmly. A consequence of this experimental arrangement is that the lattices decrease in thickness but not in lateral area \citep{Grinnell1984,Guidry1985}. The tension in such lattices is distributed anisotropically, while fibroblasts develop an elongated bipolar appearance, orienting themselves along the lines of tension \citep{Stopak1982,Bellows1982,Grinnell1994,Tamariz2002}. This reorganisation causes collagen fibrils to become oriented in the same plane as the substrate, which in turn gives rise to mechanical loading within the matrix. The contraction of such lattices, which involves a reduction in thickness alone, gives rise to a mechanically stressed tissue resembling granulation tissue, and it has therefore been proposed that such lattices can be used to model the early stage of wound healing when the granulation tissue begins to develop and exert stresses on its environment \citep{Grinnell1994}. The rate and extent of contraction of the lattice is similar to that in experiments where fibroblasts are embedded within such lattices~ \citep{Grinnell1984,Guidry1985,Guidry1986}. Fibroblasts in such lattices organise a fibronectin matrix and develop prominent actin stress fibres \citep{Farsi1984,Mochitate1991,Halliday1995}, which indicate  that some fibroblasts have differentiated into the protomyofibroblast phenotype. It has also been observed that fibroblasts in restrained matrices develop fibronexus junctions \citep{Tomasek2002}, which allow TGF-$\beta$, if present, to further stimulate the differentiation of protomyofibroblasts into the myofibroblast phenotype \citep{Arora1999,Vaughn2000}.

A \textit{Stress-relaxed FPCL} is prepared by polymerising collagen lattice, allowing it to attach to a tissue culture dish for a set period of time and then detaching it \citep{Tomasek1992}. In such lattices, tensile stress develops while the matrix is anchored and these stresses are relieved via a sudden smooth muscle-like contraction when the matrix is released, as the cell extensions collapse and the stress fibres disappear \citep{Mochitate1991,Tomasek1992,Lin1997}. It has been proposed that stress-relaxed FPCLs can be used to model scar tissue, or the transition from granulation tissue to replacement dermis in wound healing or in tissue repair \citep{Carlson2004}.

As discussed earlier, the contraction of an FPCL results from a fundamental reorganization of the lattice structure, and is hence effectively ``permanent''. We now discuss the mechanisms that have been proposed to explain the contraction of FPCLs.

\subsection{Mechanical properties}
\label{S:FPCLpermcontract}

It is known that the degradation and replacement of collagen fibres is not a significant mechanism of contraction in attached lattices. This was demonstrated experimentally using radiolabelled collagen, where it was found that the concentration of proteins in the gel remains constant during the reorganisation process~\citep{Guidry1985}. Indeed, it seems plausible to assume that contraction in all types of FPCLs is largely a result of the rearrangement of pre-existing collagen fibres by fibroblasts. Even still, the precise mechanism of lattice contraction is thought to vary, depending on the density and mechanical state of the lattice.

The prime mechanism of stressed lattice contraction is believed to be \textit{cell contraction}, which is associated with the protomyofibroblast and myofibroblast phenotypes that are prevalent in such environments~\citep{Dallon2008}. When these cells contract, they pull on the surrounding lattice, causing it to contract with them. It has been suggested that contraction in free-floating lattices of moderate cell density occurs through a \textit{cell traction} mechanism~\citep{Dallon2008}, in which cell locomotion results in the compacting of collagen fibres by bundling thin fibrils~\citep{Harris1980,Ehrlich2003}. As discussed in~\citet{Dallon2008}, the contraction of free-floating lattices of high cell density is believed to occur through a \textit{cell elongation and spreading} mechanism, in which fibroblasts pull collagen fibrils towards them, thus compacting the gel.

\begin{figure}[ht]
\center{\includegraphics[width=0.45\textwidth]{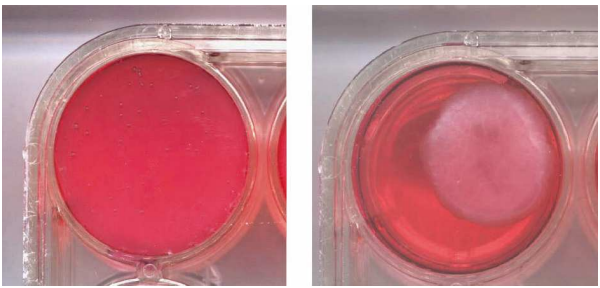}}
\caption{Contraction of a solidified free-floating FPCL. The left image shows the initial configuration of a collagen gel, while the right image shows the configuration of the same gel after 48 hrs. (With kind permission from Springer Science+Business Media: \emph{Methods in Molecular Biology: Kidney Research}, \textit{Chapter 14. Cell-Populated Floating Collagen Lattices: An In Vitro Model of Parenchymal Contraction}, 466, 2009, 1-11, Kelynack, K.~J., Figure 14.1).}
\label{fig:FPCL}
\end{figure}

A salient feature of FPCLs is that they provide a testbed for investigating the how the interplay between fibroblasts and the collagen matrix can cause the latter to permanently contract. Under the action of fibroblasts, the density of neighbouring fibrils increases locally and consequently the volume of the collagen lattice decreases~\citep{Grinnell2003}. An example of such behaviour, taken from an experiment by~\citet{Kelynack2009}, is shown in Fig.~\ref{fig:FPCL}.

The observation that the lattices only re-expand partially crucially indicates that contraction is not simply the elastic response of the collagen lattice to traction forces applied by cells. Unlike other forms of remodelling, such as the hardening of the human eye lens described by \citet{Augusteyn2010}, which could be modelled using a time-dependent stress-strain relationship,  FPCL contraction involves some form of cell-induced `plastic' behaviour. That is, the active lattice remodelling by cells causes the unloaded state of the lattice to change over time in a manner analogous to classical plasticity. Hence, a central aim of this paper is to develop a mathematical framework for FPCL contraction that takes into account the evolving unloaded state. Our approach quantitatively captures key features of the contraction process, which would be impossible to achieve with, for example, a Kelvin-Voigt viscoelastic constitutive law. We now briefly outline some of the previous mathematical approaches to modelling the behaviour of FPCLs.

\subsection{Modelling FPCL contraction}
\label{S:PrevModels}

Most previous models of FPCL contraction have incorporated a viscoelastic framework. The earliest such model was developed by~\citet{Moon1993}, who adapted the~\citet{Tranquillo1992} model of dermal wound healing to describe the contraction of a collagen microsphere. This model could not, however, account for the permanence of matrix contraction. An alternative framework, which addresses this issue, but which is only valid for small displacement gradients, was developed by~\citet{Barocas1995} who replaced the Kelvin-Voigt constitutive law in the Moon-Tranquillo model with a Maxwell constitutive law. In contrast,~\citet{Ferrenq1997} retained the linear Kelvin-Voigt constitutive law, but restricted their focus to situations in which the displacement gradient is small and linear theory is valid. A different approach, that used the theory of mixtures to describe the interaction between the fibrous lattice and the permeating fluid medium, was the biphasic model of collagen lattice contraction developed by Barocas and coworkers~\citep{Barocas1994,Barocas1997}. This model takes into account several important effects, including the partial expansion of collagen lattices after cell traction stresses are removed. Subsequently, Barocas and coworkers have used similar approaches to describe several experiments on collagen lattices (see, for example,~\citet{Chandran2004,Knapp1999,Schreiber2003}), using models that include fibroblast traction as an additive contribution to the total stress. Recent approaches to modelling FPCL contraction include models by Marquez, Zahalak and coworkers (for instance~\citep{Marquez2005b,Pryse2003,Zahalak2000}), who considered individual cells and used Eshelby's solution to describe the local strain fields and \citet{Green2013}, who considered spherical-symmetric collagen lattices and treated the gel as a compressible Stokes fluid.

While these models have significantly advanced the understanding of this process, there have not yet been any models that explicitly take into account the continual mechanical restructuring of such lattices. In the next section we introduce a morphoelastic model, based on the concept of an ``effective strain'' that can be used to capture the remodelling of FPCLs.


\section{A morphoelastic model for the contraction of fibroblast-populated collagen lattices}
\label{S:application}

In this section we develop a mechanical model that can be used to describe the contraction of a collagen lattice by fibroblasts.
As described in standard continuum mechanics texts (for example \citet{GonzalezCM}), it is common to work in either a `Lagrangian' (or `material') coordinate system, in which each particle is labelled according to its position in the initial configuration of the body, or in an `Eulerian' (or `spatial') coordinate system, in which each particle is labelled according to its current position. Holding the Lagrangian coordinate constant corresponds to observing a single particle, while holding the Eulerian coordinate constant corresponds to observing a single point in space.

As we discuss in Sec.~\ref{SS:extendedmodel}, it is often appropriate to treat contracting FPCLs as $1$-D bodies. Indeed, \citet{Ferrenq1997} constructed a model of FPCL contraction in $1$-D Cartesian coordinates and various authors have assumed radial symmetry to develop $1$-D descriptions of FPCL contraction \citep{Moon1993,Ramtani2002,Ramtani2004}. We hence derive a one-dimensional model, based on the following equation (derived explicitly in Appendix~\ref{S:Theory}) that is appropriate for describing the mechanical behaviour of morphoelastic solids with small effective strain
\begin{gather}
\label{eq:morpho1}
\pdiff{e^E}{t} + \pdiff{}{x} \left(e^E \, v \right)  =  \pdiff{v}{x} - g(x,\,t)\,,
\end{gather}
where $e^E$ is the Eulerian strain, $v$ is the velocity of a point in the material and $g(x,\,t)$ describes the rate of growth. As discussed in more detail in Appendix~\ref{S:Theory}, Eulerian strain is taken here to be the \textit{effective} Eulerian strain, a local, dimensionless measure of the difference between the current state and the zero stress state, while the rate of growth is a measure of the rate at which the zero stress state becomes larger over time. In Sec.~\ref{SS:cellcontract}, we take advantage of the simplifications that can be made to eq.~(\ref{eq:morpho1}) when the deformation of the lattice might be large, but the difference between the current state and the stress-free state is always small. In particular, we introduce a form for the growth function, $g(x,\,t)$, based on a plausible mechanism for the contraction of the lattice by fibroblasts. This ultimately leads to a full model of FPCL contraction, which we present in Sec.~\ref{SS:extendedmodel}.


\subsection{A cell-based contraction model}
\label{SS:cellcontract}

In the experiments performed by \citet{Guidry1985}, lattices were contracted by fibroblasts and then allowed to undergo a partial re-expansion after reorganisation is inhibited. \textcolor{black}{This re-expansion was faster than the contraction process, but far from instantaneous,} indicating that the time scale associated with the viscous relaxation of collagen lattices should not be ignored. It is hence preferable for us to use a Kelvin-Voigt viscoelastic constitutive law to relate stress and effective strain instead of the purely elastic law, namely $\sigma = E \, e^E$, described in Appendix~\ref{SS:Effstrain}.

Additionally, the large deformations associated with lattice contraction may cause significant changes to the elastic properties of the collagen: We expect the collagen to become stiffer as it becomes denser. Following \citet{Ramtani2002} and \citet{Ramtani2004}, we therefore propose that the elastic modulus of the collagen should be a function of collagen density. Incorporating viscoelasticity and the changing elastic modulus (but ignoring the activity of cells), this means that an appropriate constitutive law for a collagen lattice will take the form
\begin{equation}
 \sigma = \mathcal{E}(\rho) \, e^E + \mu \, \pdiff{v}{x}, \label{viscoelasticconst}
\end{equation}
where $\mathcal{E}(\rho)$ is the elastic modulus, $\rho$ is the collagen density and $\mu$ is the collagen viscosity.

Next, we develop an expression for $g(x,\,t)$ that captures how cells actively rearrange the fibres of the collagen lattice by applying traction -- a microscopic mechanism that causes a modification in size of the zero stress state. Our approach is to imagine the collagen lattice as a $1$-D body containing evenly-spaced fibroblasts, each of which effectively acts as a force dipole by pulling on the collagen lying on either side of it and compressing the collagen directly under it. We assume that each fibroblast rearranges the collagen directly under itself, thus evolving the lattice to a permanently compressed state. Thus, $g(x,\,t)$ will be negative and directly proportional to the compressive strain in the lattice under the cells. Specifically, we assume that $g(x,\,t)$ is proportional to the amount that the strain in the region covered by the cells $e_\text{cells}$ exceeds (\emph{i.e.}~is more negative than) a critical level of contraction, $-\hat{e}_\text{crit}$.

\textcolor{black}{It has been observed that even though fibroblasts lead to substantial deformation of the environment, large strains are confined to regions around each cell~\citep{Sander2013}, which results in heterogeneity in collagen density. Hence, to realistically describe the interplay of cells with their environment, one would need to assume that gel compaction is not spatially homogeneous, along the lines of the models by~\citet{Evans2009} or \citet{Stevenson2010}. This approach would introduce a range of complications that do not appear to be physically relevant to FPCLs; as noted in \cite{Grinnell1984}, apparently homogeneous compaction of a gel is observed even when cells are cultured purely on the gel surface.}

\begin{figure}[t]
\centering
{\includegraphics[width=0.45\textwidth]{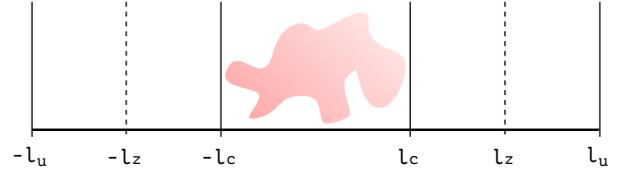}}
\caption{A single cell within a periodic unit. The cell extends from $-l_c$ to $l_c$ and the unit extends from $-l_u$ to $l_u$. The entire region has a uniform zero stress state and the length of the unit at zero stress is $l_z$.
}
\label{F:CellDiagram}
\end{figure}

\textcolor{black}{For simplicity and brevity, in the analysis that follows we assume} that the FPCL consists of a periodic array of identical units, each of which contains a single cell at its centre (see Figure \ref{F:CellDiagram}), and hence that the fibroblast density is spatially constant. \textcolor{black}{We assume that the traction forces applied by cells create regions of the lattice (under the cells) that are more compressed than other regions of the lattice (outside the cells). However, we also assume that the zero stress state remains homogeneous in space even as it evolves over time. This modelling choice is made in order to account for some effects of cell mobility. We assume that cells make local changes to the zero stress state, and then move nearby and repeat this process. The overall effect of this will be for the changes to the zero stress state to be homogenised through space.}

\textcolor{black}{Using our assumption of a periodic array of cells}, we construct a local coordinate system in which a given unit extends from $-l_u$ to $l_u$ and the cell extends from $-l_c$ to $l_c$. Thus, the cell density, $n$, is proportional to $l_c/l_u$. We also assume that the zero stress state is uniform across the unit, and that if the unit was stress-free throughout it would extend from $-l_{z}$ to $l_{z}$. Hence, the average effective strain throughout the unit is
\begin{equation}
 e_\text{avg} = \frac{l_u-l_z}{l_u}. \label{eavgdefn}
\end{equation}
Note that \textcolor{black}{a further advantage of the assumption of uniform cell density across the array is that} this expression for strain will be equivalent to that measured in a representative volume element containing many cells. We shall therefore use this measure of strain in our macroscopic model.

As the cell in each unit pulls the lattice in towards itself, the collagen under the cell will be under less tension (or more compression) than that over the rest of the unit. More formally, a cell in a given unit can be treated as a pair of body forces, both of magnitude $\tau_c$; at $-l_c$ the cell pulls the collagen lattice to the right, while at $l_c$ the cell pulls the lattice to the left. Neglecting inertial terms, the momentum balance equation within a single unit is then
\[
 \pdiff{\sigma_u}{x} = \tau_c \, \left(\delta(x - l_c) - \delta(x+ l_c)\right),
\]
where $\sigma_u$ is the local stress and $\delta$ is the Dirac delta function. Integrating this equation with periodic boundary conditions, we find that the stress within the unit is
\begin{equation*}
 \sigma_u = \begin{cases}
             \sigma_\text{out}, & l_c < |x| < l_u, \\[6pt]
             \sigma_\text{out} - \tau_c, & |x| < l_c.
            \end{cases}
\end{equation*}
where $\sigma_\text{out}$ represents the stress in the regions outside the cells, which we assume to be governed by the constitutive law \eqref{viscoelasticconst}, so that
\begin{equation}
 \sigma_\text{out} = \mathcal{E}(\rho) \, e_\text{out} + \mu\,\pdiff{v}{x}\,. \label{viscoelastoutside}
\end{equation}
Since each cell must apply a balanced pair of body forces to the lattice, it follows that either the boundary of the lattice occurs in a region where $\sigma = \sigma_\text{out}$, or there is an additional surface traction term that accounts for the effects of the end of the cell. In either case, we find that it is most convenient to express macroscopic boundary conditions in terms of $\sigma_\text{out}$: a free-floating lattice will have $\sigma_\text{out} = 0$ on the boundary, while a spring connected to the boundary (as in, for example, \citet{Marenzana2006}) should be expressed as a relationship between $\sigma_\text{out}$ and the displacement at the boundary.

From the definition of strain above, we can express $\tau_c$ as
\begin{equation}
 \tau_c = \sigma_\text{out} - \sigma_\text{cells} \,, \label{tauc_straindff}
\end{equation}
where $\sigma_\text{cells}$ is the stress in the region covered by the cell. Assuming that viscous stresses are uniform (or negligible) between $l_c$ and $-l_c$, and using a constitutive law for $\sigma_\text{cells}$, we obtain the expression
\begin{equation*}
 e_\text{cells} = \frac{\sigma_\text{out} - \tau_c}{\mathcal{E}(\rho)}\,.  
\end{equation*}
As mentioned earlier, we assume that the growth rate $g(x,\,t)$ is dependent on the physical contraction experienced by the lattice directly under the cells and is proportional to $- (e_\text{cells} - \hat{e}_\text{crit})$. Moreover, the rate of contraction is proportional to the cell density, $n$, as this is representative of the proportion of the lattice that is accessible to the fibroblasts. Thus, we obtain the following constitutive law for $g(x,\,t)$:
\begin{equation}
 g(x,\,t) = - \theta\, n\, \left(- \frac{\sigma_\text{out} - \tau_c}{\mathcal{E}(\rho)} - \hat{e}_\text{crit}\right)^{+}\,, \label{growthrate}
\end{equation}
where $\theta$ is a constant of proportionality with dimensions of cell density$^{-1}$ time$^{-1}$ and the positive part operator $(X)^{+}=X$ for any positive $X$ and is zero otherwise. As the condition
\[
\hat{e}_\text{crit} < - \frac{\sigma_\text{out} - \tau_c}{\mathcal{E}(\rho)} 
\]
is always satisfied in most practical situations, we neglect the positive part operator in \eqref{growthrate} for the rest of our analysis.

Next, we note that since the zero stress state is uniform throughout the unit,
\begin{equation}
 l_u - l_z = e_\text{cells} \, l_c + e_\text{out} \, (l_u - l_c)\,. \label{ecellsandout}
\end{equation}
Combining this with \eqref{eavgdefn} yields the expression
\[
 e_\text{avg} = e_\text{out} + \frac{l_c}{l_u} \big(e_\text{cells} - e_\text{out}\big).
\]
Using \eqref{tauc_straindff}, and defining $\sigma_c$ as rescaling of $\tau_c$ so that $n \, \sigma_c = l_c/l_u \, \tau_c$, we obtain
\[
 e_\text{out} = e_\text{avg} + \frac{n \, \sigma_c}{\mathcal{E}(\rho)}\,.
\]
Substituting into \eqref{viscoelastoutside}, gives us our constitutive relationship between stress and strain:
\begin{equation*}
 \sigma_\text{out} = \mathcal{E}(\rho) \, e_\text{avg} + \mu\,\pdiff{v}{x} + n \, \sigma_c\,. \label{cellelasticconst}
\end{equation*}

Note that the $n \, \sigma_c$ term is analogous to the cell traction stress term found in other models of lattice contraction and dermal wound healing \citep{Moon1993,Tranquillo1992,Ferrenq1997,Tracqui1995}. Indeed, by making $\sigma_c$ a function of the lattice density and/or cell density (to incorporate the effects of crowding on traction stress, for example) we can recover identical expressions to those used in these earlier papers. However, our approach differs in that traction stress emerges naturally from the assumption that cells apply body forces to the collagen lattice, rather than by requiring that this stress term be incorporated into the constitutive law. Nevertheless, the consistency between these two treatments of cell traction gives us further confidence that both approaches are valid and useful.

Dropping the subscripts on $e_\text{avg}$ and $\sigma_\text{out}$, we obtain the following set of equations to describe the mechanical behaviour of a contracting viscoelastic lattice:
\begin{subequations}
\label{eq:1Dcellcontract}
\begin{gather}
  \label{eq:strain}
  \pdiff{e}{t} + \pdiff{}{x} \left(e\,v\right) = \pdiff{v}{x} + \theta\, n \left(-\frac{\sigma - n_{l} \, \sigma_{c}}{\mathcal{E}(\rho)} - \hat{e}_\text{crit}\right)\,, \\[6pt]
  \label{eq:stress}
  \sigma = \mathcal{E}(\rho) \, e + \mu\,\pdiff{v}{x} + n \, \sigma_c\,, \\[6pt]
  \label{eq:forcebalance}
  \pdiff{\sigma}{x} = 0\,,
\end{gather}
\end{subequations}
where $n_{l} = n \, l_u / l_c$ is the constant of proportionality that relates the cell density, $n$, to the cell length per unit length, $l_c / l_u$.

A full description of the mechanics of a contracting lattice will also require an initial condition on $e$ and two boundary conditions, either on $\sigma$ or $v$, that specify whether the ends of the lattice are tethered or stress-free. We now combine \eqref{eq:1Dcellcontract} with some assumptions about fibroblast motion and interactions in FPCLs to construct a full morphoelastic model of FPCL contraction.

\subsection{The morphoelastic model of FPCL contraction}
\label{SS:extendedmodel}

In the following we assume that the contraction of the lattice is due to the rearrangement of collagen fibrils by fibroblasts alone, with the expectation that the results obtained using this assumption will be quantitatively similar to those obtained using a model that takes into account the r\^{o}le of protomyofibroblasts. Additionally, we assume that TGF-$\beta$ is not present in the collagen lattice. As protomyofibroblasts require the presence of TGF-$\beta$ to differentiate into myofibroblasts \citep{Tomasek2002}, we hence do not consider the activity of the latter. Furthermore, while it has been observed that fibroblasts in free-floating lattices become quiescent \citep{Rosenfeldt2000} and a small fraction of these cells undergo apoptosis (cell death) \citep{Fluck1998,Grinnell1999}, we shall neglect these processes in our model.

Experimental data for the contraction of FPCLs are typically obtained by measuring the changing diameter (in the case of free-floating FPCLs) or thickness (in the case of attached FPCLs) of the lattice. The behaviour of such lattices can hence be modelled by considering one spatial dimension, that represents either the radius or thickness of the lattice depending on the type of FPCL being considered. It should be noted that the mechanical model presented in Appendix~\ref{S:Theory} assumes a single Cartesian spatial dimension and cannot be easily modified to obtain equations for the mechanics of radially symmetric lattice contraction. As a result, our model is best suited to attached, rather than free-floating, FPCLs. Nevertheless, as seen later in this section, our model is equally successful at fitting data obtained from both these lattice types.

\begin{figure*}[ht]
\center{
\includegraphics[width=0.75\textwidth]{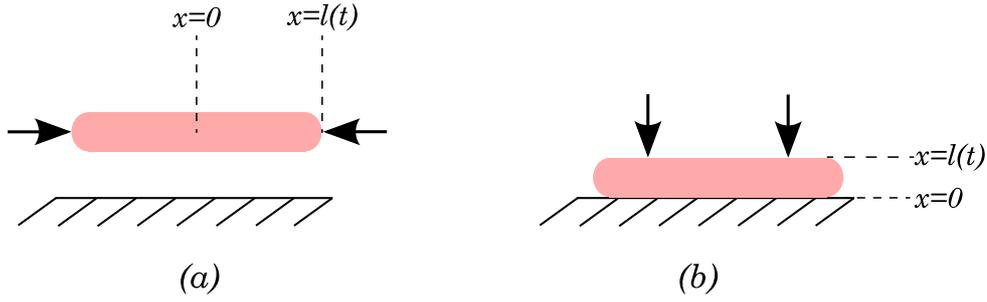}
}
\caption{Sketch of the different situations that can be described using our morphoelastic model. Here the dashed lines represent the substrate and the arrows represent the directions of contraction of the collagen lattice (pink). (a) In the case of a free-floating FPCL, we have symmetry about $x=0$. (b) In the case of an attached FPCL, $x=0$ represents the base of the lattice and $x=l(t)$ is the contracting edge. This situation can be thought of as the case of a lattice with a spring of infinite stiffness attached at $x=0$.}
\label{fig:spring}
\end{figure*}

Following the conventions used in Appendix~\ref{SS:MultiDecomp}, we use $x(X,\,t)$ to represent the position at time $t$ of a particle initially located at $X$ and $X(x,\,t)$ to represent the initial position of a particle at position $x$ at time $t$. Since the lattice changes in size over time, the domain of interest can be expressed as $0 \leq x \leq l(t)$ or, equivalently, $0 \leq X \leq l_0$, where $l_0 = l(0)$. As illustrated in Fig.~\ref{fig:spring}, the point $x = 0$ is taken to represent either the centre of a floating lattice or the fixed tethering point of an attached lattice. In either case, this point is fixed and we hence find that $x(0,\,t) = X(0,\,t) = 0$.

We next define the Eulerian displacement, $u(x,\,t) = x - X(x,\,t)$, and the Eulerian displacement gradient, $w(x,\,t) = {\partial u}/{\partial x}$. Following the definitions of $x$ and $X$, it follows that $u(x,\,0) = w(x,\,0) = 0$. Moreover, the fact that the centre or tethering point is fixed implies that $u(0,\,t) = 0$. Another important kinematic variable is the velocity, which is defined as
\begin{equation}
v(x,\,t) = \frac{\rm{D}u}{\rm{D}t} = \frac{\partial u}{\partial t} + v \, \frac{\partial u}{\partial x}\,,
\label{eq:velocity}
\end{equation}
with $v(0,\,t) = 0$ at the fixed centre/tethering point.

Next, we introduce expressions for the densities of the ECM, $\rho(x,\,t)$, and the fibroblast cells, $n(x,\,t)$. Since \citet{Guidry1985} observed that collagen synthesis and degradation are insignificant in FPCLs, we propose that contraction proceeds only by rearrangement of the collagen lattice. In other words, we assume that the net creation of each species is zero during the timescale of the experiment. Furthermore, for this study, we assume that there is no migration of fibroblasts. As both species are subjected to passive advection, this assumption allows us to use simple continuity equations for each of the species, which can be solved explicitly \citep{Clement1978} to give
\begin{eqnarray}
\label{eq:n}
n(x,\,t)&=&n_{0}\,(1-w(x,\,t))\,,\\
\label{eq:rho}
\rho(x,\,t)&=&\rho_{0}\,(1-w(x,\,t))\,,
\end{eqnarray}
where $n_{0}$ and $\rho_{0}$ are the (spatially uniform) initial densities of fibroblast cells and the ECM, respectively.

We assume that the lattice is initially relaxed ($e(x,\,0)=0$), and that lattice stiffness increases with density, according to the simple linear relationship
\[
 \mathcal{E}(\rho)= \mathcal{E}_0 + k \, \frac{\rho - \rho_0}{\rho_0}\,,
\]
where $\mathcal{E}_0$ is the elastic modulus when $\rho = \rho_0$ and $k$ is a positive constant. It should be noted that other forms have been proposed to describe the dependence of the elastic modulus on density. For example, Ramtani and coworkers describe this relationship as a power law \citep{Ramtani2002,Ramtani2004}. In the absence of further data, however, we use a simple, linear law.

The case of a free-floating FPCL (Fig.~\ref{fig:spring}a) or an attached FPCL (Fig.~\ref{fig:spring}b) can be described by using the stress-free boundary condition
\begin{equation}
\sigma(l(t),t)=0\,,
\label{eq:stressboundary}
\end{equation}
and the displacement at $x = l(t)$ can be obtained by noting that
\[\int^{l(t)}_{0}\,w(\xi,t) \mathrm{d}\xi = u(l(t),t) = l(t)-l_{0}\,.\]
As we shall see, $w(x,\,t)$ is spatially homogeneous in the problem that we analyse. In this situation, it follows that
\begin{equation}
l(t)=\frac{l_{0}}{1-w}\,.
\label{eq:Lt}
\end{equation}
Taking the spatial derivative of \eqref{eq:velocity} and rearranging, we obtain the following evolution equation for $w(x,\,t)$:
\begin{equation}
\pdiff{w}{t}+\pdiff{}{x}\left(w\,v\right)=\pdiff{v}{x}\,.
\label{eq:weqn}
\end{equation}
In conjunction with the equations for the evolution of the effective strain (\ref{eq:strain}), the constitutive law for the total stress (\ref{eq:stress}) and the force balance equation (\ref{eq:forcebalance}), equations (\ref{eq:n})-(\ref{eq:weqn}) comprise an Eulerian model for the contraction of collagen by fibroblasts. This system can be non-dimensionalised by introducing the dimensionless variables
\begin{align*}
&x^{*}=\frac{x}{l_{0}}\,,&
&l^{*}=\frac{l}{l_{0}}\,,&
&t^{*}=\frac{t}{h}\,,&
&v^{*}=\frac{v\,h}{l_{0}}\,,&
&w^{*}=w\,,\\
&e^{*}=e\,,&
&n^{*}=\frac{n}{n_{0}}\,,&
&\rho^{*}=\frac{\rho}{\rho_{0}}\,,&
&\sigma^{*}=\frac{\sigma}{\mathcal{E}_{0}}\,,&
&\mathcal{E}^{*}=\frac{\mathcal{E}}{\mathcal{E}_{0}}\,,&
\end{align*}
and the dimensionless constants
\begin{equation}
\begin{aligned}
&\bar{\mu} = \frac{\mu}{\mathcal{E}_{0}\,h}\,,&
&\bar{k} = \frac{k}{\mathcal{E}_{0}}\,,&
&\bar{\tau} = \frac{\sigma_{c} \, n_0}{\mathcal{E}_{0}}\,,\\
&\bar{\theta} = h\,\theta\,n_{l}\,,&
&\bar{e}_{\rm crit} = \frac{\hat{e}_{\rm crit}\,n_{0}}{n_{l}}\,,&
\end{aligned}
\label{eq:parameters}
\end{equation}
where, for ease of comparison with experiment (see later), we choose $h=1$ hr.

\subsection{Lagrangian description of the model}

As this Eulerian model has a moving boundary at $x = l(t)$, we now adopt a Lagrangian coordinate system to obtain a model for a system in which the right hand boundary is fixed. Conveniently, we find that transforming our equations to Lagrangian variables is equivalent to converting to characteristic variables: we ultimately obtain a system where the only differentiation is with respect to time. In the derivation that follows, it is important to distinguish between partial time derivatives with the Eulerian spatial coordinate held fixed and partial time derivatives with the Lagrangian spatial coordinate  $X$ held fixed. Hence, we introduce the `Lagrangian time variable', $T$, and we transform the entire problem from $(x,\,t)$ coordinates to $(X,\,T)$ coordinates.

\begin{center}
\begin{table*}[t]
\caption{Summary of the experimental values for the number of fibroblast cells in the collagen gel and the initial gel diameter, as well as the estimated values of the initial cell traction stress $\sigma_0$ and the parameters $\bar{\mu}$, $\bar{\tau}$ and $\bar{e}_{\rm crit}$, assuming $\mathcal{E}_{0} = 1000$ Pa and $F_\text{cell}=500n$N/cell. The remaining model parameters $\bar{\theta}$ and $\bar{k}$ are optimized to fit the model to each set of experimental data (see Table~\ref{tab:summary2}).}
\centering
\begin{tabular}{|c|c|c|c|c|c|c|c|}
\hline
{Experiment}  &{C (cells)}  &{d (cm)}  &{$\sigma_0$ (Pa)}  &{$\bar{\mu}$}  &{$\bar{\tau}$}  &{$\bar{e}_{\rm crit} (\times 10^{-3})$}\\
\hline
{Bell et al. (1979)}        &{$7.5\times 10^{6}$} &{$5.3$}  &{$1700$}  &{$2.78$} &{$1.699$}     &{$16.99$}\\
{Talas et al. (1997)}       &{$2.5\times 10^{5}$} &{$2.18$} &{$334.9$} &{$2.78$} &{$0.3349$}     &{$3.35$}\\
{Feng et al. (2003)}        &{$2.5\times 10^{6}$} &{$10$}   &{$159.2$} &{$2.78$} &{$0.1592$}     &{$1.59$}\\
{Guidry \& Grinnell (1985)} &{$1.0\times 10^{5}$} &{$1.2$}  &{$442.1$} &{$2.78$} &{$0.4421$}     &{$4.42$}\\
\hline
\end{tabular}
\label{tab:summary}
\end{table*}
\end{center}

Firstly, we introduce Lagrangian displacement $U$ and Lagrangian velocity $V$, which are defined as
\[
U(X,\,T) = x(X,\,T) - X,\;\;\;\; V(X,\,T) = \frac{\partial U}{\partial T}\,.
\]
These are both identical to $u(x,\,t)$ and $v(x,\,t)$ with just a transformation of the independent variables (see Appendix~\ref{SS:Effstrain} for details). Now, we define the Lagrangian displacement gradient, $W(X,\,T) = {\partial U}/{\partial X}$, which is different from $w$ since differentiation with respect to $x$ is different from that with respect to $X$. As described in Appendix~\ref{sec:AppTransf}, however, the definitions of $W$ and $V$ lead to the identities
\begin{eqnarray}
\label{eq:LEx}
\pdiff{}{x}&\equiv&\frac{1}{1+W}\pdiff{}{X}\,,\\
\label{eq:LEt}
\pdiff{}{t}&\equiv&\pdiff{}{T}-\frac{V}{1+W}\pdiff{}{X}\,,
\end{eqnarray}
and hence $w = W/(1+W)$. Based on these identities, it follows that
\begin{equation}
\label{eq:advectivederiv}
\pdiff{\phi}{t}+\pdiff{}{x}\left(\phi\,v\right)=\frac{1}{1+W}\pdiff{\Phi}{T}\,,
\end{equation}
where $\Phi = \phi \, (1 + W)$ and where $\phi$ is a general scalar quantity. This motivates the introduction of Lagrangian variables $N = n \, (1 + W)$, $R = \rho \, (1 + W)$, $S = \sigma \, (1 + W)$ and $E = e \, (1 + W)$, so that all of the advective time derivatives simply become partial time derivatives. From (\ref{eq:n}) and (\ref{eq:rho}), we obtain the following exact expressions for the scaled densities $N$ and $R$ in the `normal' contracting situation
\[N(X,\,T)\equiv 1\,,\qquad R(X,\,T)\equiv 1\,,\]
Note that the situation in which lattice reorganisation is inhibited by the addition of cytochalasin D at some time, $T_{\rm inh}$, can be described by taking
\begin{equation}
N(X,T)= \left\{
\begin{array}{cc}
1\,,&\qquad T\leq T_{\rm inh}\,,\\
0\,,&\qquad T>T_{\rm inh}\,.
\end{array}\right.
\label{eq:N_inh}
\end{equation}
From (\ref{eq:weqn}), (\ref{eq:strain}), (\ref{eq:stress}) and (\ref{eq:forcebalance}) we obtain the equations
\begin{eqnarray*}
\pdiff{W}{T}&=& \pdiff{V}{X}\,,\\
\pdiff{E}{T}&=& \pdiff{V}{X} + N\,\bar{\theta}\,\left(-\frac{\frac{n_{0}}{n_{l}}\frac{S}{1+W} - \bar{\tau}}{\mathscr{E}(W)} - \bar{e}_{\rm crit}\right)\,,\\
S(X,\,T) &=& \mathscr{E}(W)\,E + \bar{\mu}\,\pdiff{V}{X} + \bar{\tau}\,N\,,\\
\pdiff{S}{X} &=& 0\,,
\label{eq:}
\end{eqnarray*}
where $\mathscr{E}(W)$ is
\begin{equation}
\label{eq:EW}
\mathscr{E}(W)=1-\bar{k}+\frac{\bar{k}}{1+W}\,.
\end{equation}
Using the stress-free boundary condition (\ref{eq:stressboundary}), it follows that
\[S(X,\,T)\equiv 0\,.\]
As $W$ is independent of $X$, it follows that $w$ is independent of $x$ and hence (\ref{eq:Lt}) is valid. Moreover, the fact that $W$ and $E$ are both independent of $X$ means that we can rewrite our system of PDEs as a pair of coupled ODEs for the displacement gradient and strain:
\begin{eqnarray}
\label{eq:W_L_non}
\frac{\mathrm{d}W}{\mathrm{d}T}&=& -\frac{1}{\bar{\mu}}\left(\mathscr{E}(W)\,E + \bar{\tau}\,N\right)\,,\\
\label{eq:E_L_non}
\frac{\mathrm{d}E}{\mathrm{d}T}&=& \frac{\mathrm{d}W}{\mathrm{d}T} + N\,\bar{\theta}\,\left(\frac{\bar{\tau}}{\mathscr{E}(W)}-{\bar{e}}_{\rm{crit}}\right)\,.
\end{eqnarray}
Although this model has five dimensionless free parameters:  $\bar{\mu}$, $\bar{k}$, $\bar{\tau}$, $\bar{e}_{\rm{crit}}$ and $\bar{\theta}$, as we show in the next section, we can fix three of these parameters using heuristic arguments, leaving us with a  two-component model with two free parameters. Finally, on non-dimensionalising (\ref{eq:Lt}) and converting to Lagrangian variables, we find
\begin{equation}
l(t)=1+W(T)\,.
\label{eq:lW}
\end{equation}
Thus, our transformed ODE model gives easy access to the evolving length of the lattice, the physical variable variable most easily observed in experiments.

\subsection{Comparison with experimental data}
\label{SS:numerics}

In the following, we numerically integrate the system (\ref{eq:W_L_non})-(\ref{eq:E_L_non}) and compare the results with previously obtained data. In particular, we consider the experiments of \citet{Bell1979}, \citet{Talas1997} and \citet{Feng2003}, which used free-floating FPCLs, as well of \citet{Guidry1985}, which was performed using an attached FPCL and in which lattice reorganisation was halted at various times. We first estimate values for the parameters $\mathcal{E}_{0}$, $F_\text{cell}$ and $\sigma_0$, which we then use to fix the parameters $\bar{\mu}$, $\bar{\tau}$ and $\bar{e}_{\rm crit}$ for each case (values listed in Table~\ref{tab:summary}).

The initial stiffness of a collagen gel $\mathcal{E}_{0}$ was measured by \citet{Knapp1997,Knapp1999} to be $1.185$~kPa. However the precise value of this quantity has been found to vary widely, depending on the extent of cross-linking~\citep{Discher2005}. As the initial stiffness of the gel was not measured in any of the experiments under consideration, we assume for the purposes of our simulations that the gel used in each case was a lightly cross-linked collagen lattice, and use the approximate value of the stiffness of such lattices~\citep{Discher2005}, namely $\mathcal{E}_{0}=1$~kPa.

Next we note that the total cell traction force will be a product of the applied contraction force per cell $F_\text{cell}$ and the number of cells in the lattice, $C$. On dividing this quantity by the cross-sectional area of the lattice, $A$, we obtain an expression for initial cell traction stress:
\begin{equation}
\sigma_0 = \frac{F_\text{cell} \, C}{A}.
\label{eq:sigma0}
\end{equation}
The values of $C$ and the initial lattice diameter $d$ (from which we can calculate $A$) for each of the four experiments are \textcolor{black}{listed} in Table~\ref{tab:summary}. Although the \textcolor{black}{value of $F_\text{cell}$ was not measured in any of the four experiments}, there have been several other experiments in which the contraction forces exerted by the cells have been determined using cell-populated lattices attached to force monitors, and a wide range of values of $F_\text{cell}$ has been reported, ranging from from $0.1$nN/cell \citep{Eastwood1996} to $1000$nN/cell \citep{Wakatsuki2000}. Here, we will use the result from \citet{Kolodney1992}, $F_{\rm cell}=500n$N/cell, which was obtained using fibroblast-populated matrices similar to those used in the experiments by \citet{Bell1979}, \citet{Talas1997}, \citet{Feng2003} and \citet{Guidry1985}.

Now, we note that equation \eqref{eq:stress} implies that the initial cell traction stress can also be expressed as
\[
\sigma_{0} = \sigma_c \, n_0.
\]
Rearranging  the equation for $\bar{\tau}$ in (\ref{eq:parameters}) and using (\ref{eq:sigma0}) with this alternative expression for $\sigma_{0}$, we find that
\begin{equation}
\bar{\tau}=\frac{F_\text{cell} \, C}{\mathcal{E}_{0}\,A}\,.
\label{eq:stiffness}
\end{equation}
Thus, the value of $\bar{\tau}$ can be estimated for each experiment, given the values of $C$ and $A$.

\begin{center}
\begin{table*}[t]
\caption{Values of the model parameters $\bar{\theta}$ and $\bar{k}$, obtained from the best fit of the results of the simulations to four different sets of experimental data. Goodness of fit tests are performed by comparing simulations performed using these parameter values with the corresponding experimental data and the resulting $\chi^2$ values for each case are displayed. In addition, we display $\chi^2$ values for the cases where one of $\bar{\theta}$ and $\bar{k}$ are scaled by factors of $\pm10\%$ (indicated by a corresponding superscript $\pm$), to characterize the robustness of the individual fits.}
\centering
\begin{tabular}{|c|c|c|c|c|c|c|c|}
\hline
{Experiment}  &{$\bar{\theta}$} &{$\bar{k}$} & {$\chi^2(\bar{\theta},\bar{k})$} & {$\chi^2(\bar{\theta},\bar{k}^{-})$} & {$\chi^2(\bar{\theta},\bar{k}^{+})$} & {$\chi^2(\bar{\theta}^{-},\bar{k})$} & {$\chi^2(\bar{\theta^{+}},\bar{k})$}\\
\hline
{Bell et al. (1979)}        &{$0.89$}  &{$106.62$}  &{0.0147} &{0.0313} &{0.0298} &{0.0244} &{0.0218}\\
{Talas et al. (1997)}       &{$5.10$}  &{$1294.92$} &{0.0037} &{0.0071} &{0.0063} &{0.0045} &{0.0044}\\
{Feng et al. (2003)}        &{$16.14$} &{$183.85$}  &{0.0337} &{0.0446} &{0.0434} &{0.0363} &{0.0356}\\
{Guidry \& Grinnell (1985)} &{$0.33$}  &{$2.97$}    &{0.0590} &{0.0704} &{0.0685} &{0.0921} &{0.0890}\\
\hline
\end{tabular}
\label{tab:summary2}
\end{table*}
\end{center}
\begin{figure*}[ht]
\center{
\includegraphics[width=0.75\textwidth]{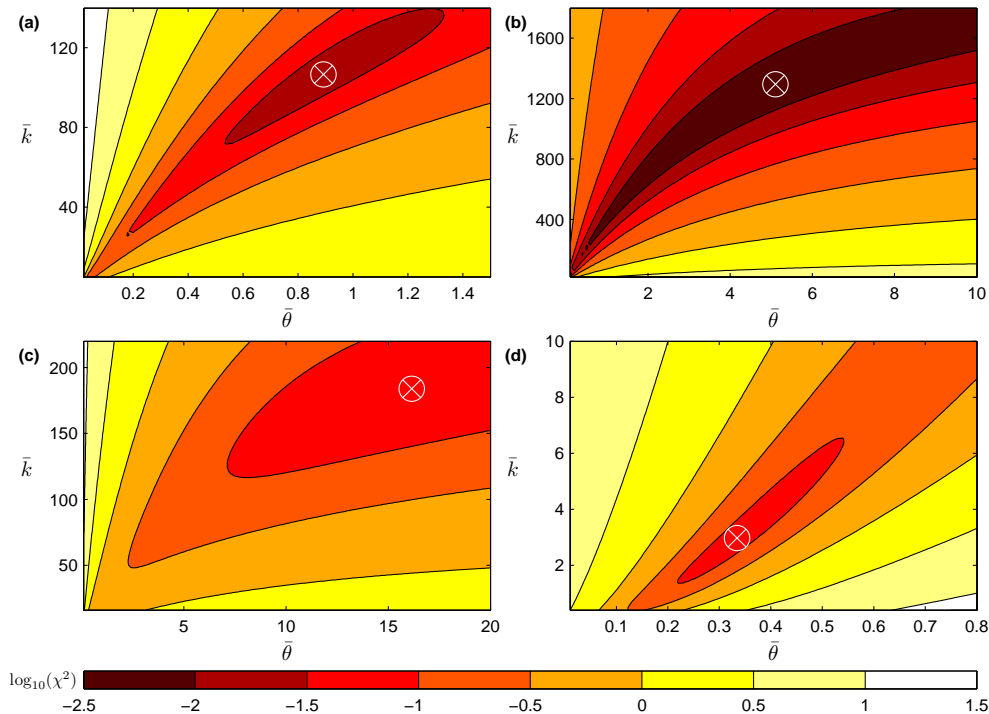}
}
\caption{
Contour plots displaying the dependence of the goodness of fit on the free parameters of the model, $\bar{\theta}$ and $\bar{k}$. To capture the variation of the goodness of fit over the parameter space, we display the logarithm (base $10$) of $\chi^2$($\bar{\theta}$, $\bar{k}$). The latter quantity is calculated by comparing results obtained by simulating the model for a given value of $\bar{\theta}$ and $\bar{k}$ with experimental data, using Eq.~(\ref{eq:chi2}). Four experimental datasets are considered, namely (a) data for lattice diameter from \citet{Bell1979}, (b) data for lattice diameter from \citet{Talas1997}, (c) data  for lattice diameter from \citet{Feng2003} and (d) data for lattice thickness from \citet{Guidry1985}. In each case, the ``best-fit'' values of $\bar{\theta}$ and $\bar{k}$ that yield the lowest $\chi^2$ are indicated by a cross within a circle.
}
\label{fig:contour}
\end{figure*}

We next note from (\ref{eq:parameters}) that the viscosity of the lattice is given by
\begin{equation}
\mu=\bar{\mu}\,\mathcal{E}_{0}\,h\,.
\label{eq:viscosity}
\end{equation}
For the purpose of our simulations, we assume that $\mu=1.0\times 10^{7}$Pa~sec, a value in agreement with \citet{Knapp1997} who measured the shear viscosity of a collagen gel to be $1.24\times 10^{7}$Pa~sec. As we assume that the initial gel stiffness is $\mathcal{E}_{0}=1$kPa, we hence  set $\bar{\mu}=2.78$ for all experiments.

Finally, we note that although $\bar{e}_{\rm crit}$ is difficult to determine experimentally, it can be observed from (\ref{eq:parameters}) that it is related to the cell density. Hence, we make the heuristic  assumption $\bar{e}_{\rm crit} = C/(C_{f}\,A)$, where $C_{f}$ represents the characteristic number of fibroblasts in an FPCL. In the following, we choose $C_{f}=2\times 10^{6}$, which is a reasonable estimate that lies within the range of values of $C$ for the experiments considered here. It is interesting to note that the above assumptions yield a simple relationship $\bar{e}_{\rm crit}=10^{-3}\,\bar{\tau}$.

The remaining parameters $\bar{\theta}$ and $\bar{k}$ are difficult to estimate from experiments. In particular, $\bar{\theta}$ arises from the constitutive law for $g(x,t)$ and there are no constraints on its value. Hence, in the following we obtain and use ``best-fit'' values of $\bar{\theta}$. Moreover, we note the power law exponent for the relationship between elastic modulus and density used Ramtani and coworkers used~\citep{Ramtani2002,Ramtani2004} is, to a first order approximation, identical to $\bar{k}$. As this exponent was taken to be a free parameter in these previous approaches, we obtain and use best-fit values of $\bar{k}$.

\begin{figure*}[ht]
\center{
\includegraphics[width=0.75\textwidth]{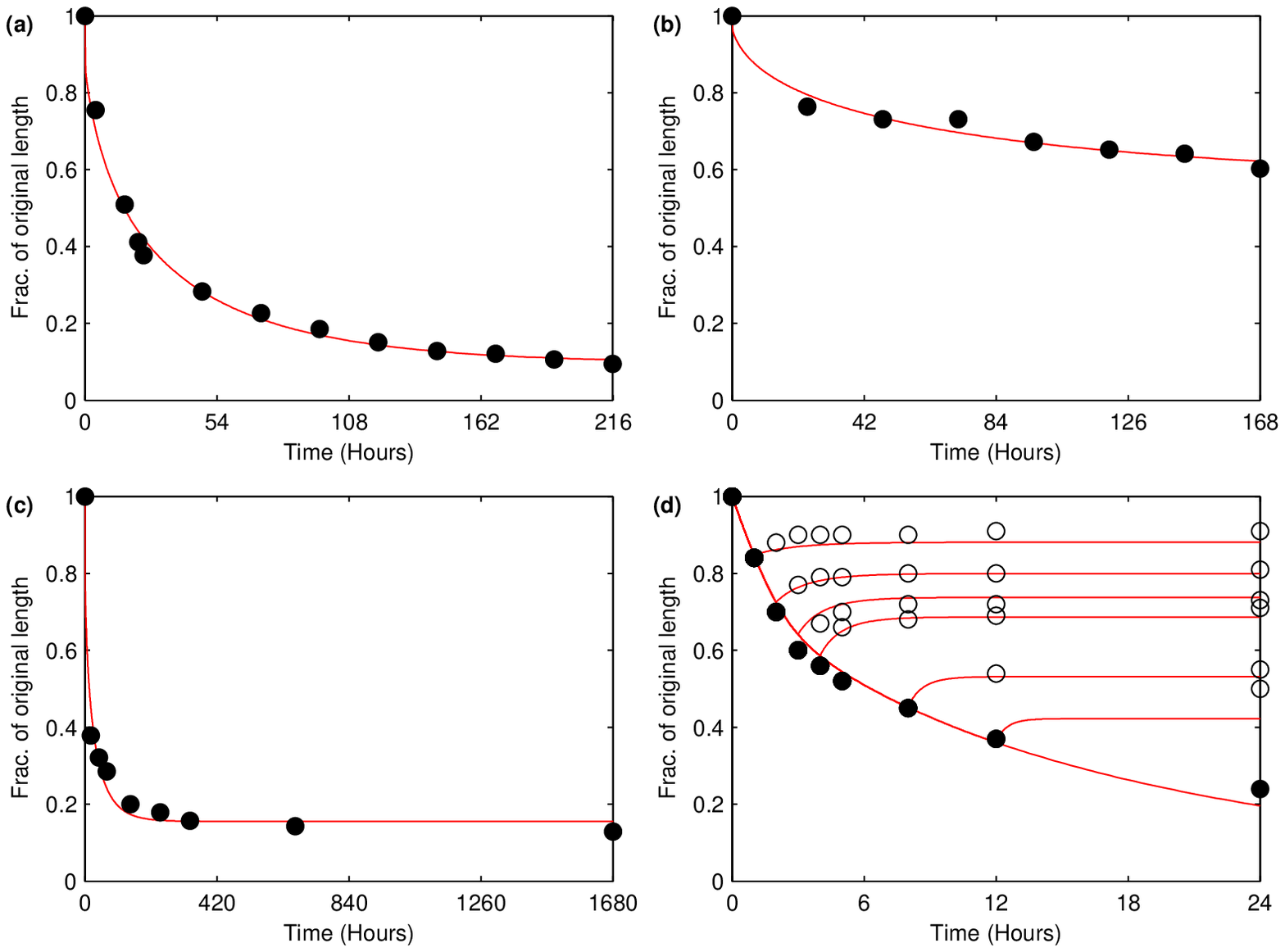}
}
\caption{Data for the fraction of the original length (diameter or thickness) of the FPCL, taken from different experiments (denoted by dark circles), superimposed with numerical results for $l(t)$  (from the expression (\ref{eq:lW}), and denoted by red solid lines) obtained by simulating the system (\ref{eq:W_L_non})-(\ref{eq:E_L_non}). The parameters $\bar{\mu}$, $\bar{\tau}$ and $\bar{e}_{\rm crit}$ are estimated heuristically from available information for each experiment, and the values used for simulations in each case are listed in Table~\ref{tab:summary}. The corresponding best-fit values of $\bar{\theta}$ and $\bar{k}$ are listed in Table~\ref{tab:summary2}. Fits are shown for (a) data for lattice diameter from \citet{Bell1979}, (b) data for lattice diameter from \citet{Talas1997}, (c) data  for lattice diameter from \citet{Feng2003} and (d) data for lattice thickness from \citet{Guidry1985}. Experimental data is available for six different cases in \citep{Guidry1985}, for which lattice reorganization is inhibited through the addition of cytochalasin D at different times $T_{\rm inh}$. We simulate this inhibition by setting $N(X,T)=0$ at $T=T_{\rm inh}$, and the results are displayed for each corresponding case. Note that the model captures the partial re-expansion of the lattice (empty circles) observed in the experiment.
}
\label{fig:fits}
\end{figure*}

We now qualitatively describe the behaviour of the FPCL in each of the four experiments under consideration by using the values displayed in Table~\ref{tab:summary} and varying the free parameters $\bar{\theta}$ and $\bar{k}$ using the MATLAB subroutine \emph{fminsearch}, such that we minimise the difference between the experimental values for the fraction of the original length (diameter or thickness) and the values of $l(t)$, given by (\ref{eq:lW}) at the corresponding measurement times. Specifically, we find the values of $\bar{\theta}$ and $\bar{k}$ that yield the lowest values of the quantity $\chi^2$, defined as:
\begin{equation}
\chi^2 = \sum\limits_{i} \frac{\left(\mathscr{O}_{i} - \mathscr{E}_{i}\right)^2}{\mathscr{E}_{i}}\,,
\label{eq:chi2}
\end{equation}
where $i$ correspond to all the experimental data points being considered, $\mathscr{E}_{i}$ are the values of $l(t)$ for these data points and $\mathscr{O}_{i}$ are the corresponding values of $1+W(T)$ obtained from the simulations. As shown in Table~\ref{tab:summary2}, we find that the $\chi^2$ values, obtained by using the optimal values of $\bar{\theta}$ and $\bar{k}$ for each case, are in the range $0.004-0.059$, indicating a good fit between theory and experiments.

To test the robustness of these fits, we determine the values of $\chi^2$ obtained by simulating our model for a wide range of choices of $\bar{\theta}$ and $\bar{k}$. In Fig.~\ref{fig:contour}, we display contour plots that indicate how $\chi^2$ varies with $\bar{\theta}$ and $\bar{k}$ for each of the four experiments mentioned earlier. In addition, for each case we display the location of the ``optimal'' choice of $\bar{\theta}$ and $\bar{k}$, obtained through the minimization procedure described above. It can be observed from this figure that small perturbations (around $\pm10\%$) in the value of either of the two parameters from the global minima will not substantially change the resulting values of $\chi^2$. Indeed, as seen in Table~\ref{tab:summary2}, the corresponding $\chi^2$ values in these cases are of a similar order of magnitude to those obtained when using the best-fit values of $\bar{\theta}$ and $\bar{k}$. This implies that the global minima displayed in Fig.~\ref{fig:contour} can be used to obtain robust fits of the simulations to each of the four datasets.

\textcolor{black}{It is important to note that our choice of the global minima of $\chi^2$ is in order to provide a unifying criterion for the selection of $\bar{\theta}$ and $\bar{k}$. Although the best-fit values listed in Table~\ref{tab:summary2} vary over a wide range, there is a large region of the ($\bar{\theta}$,$\bar{k}$)-space over which we could obtain fits that are essentially as good (see Fig.~\ref{fig:contour}). Thus, one could have, in principle, chosen very similar values of $\bar{\theta}$ and $\bar{k}$ for each of the three free-floating FPCL experiments and have still obtained good fits in each case. While such a choice might potentially be more realistic, as one would not expect much variance of these parameters across different free-floating FPCLs, it is difficult to justify the use of any pair of ($\bar{\theta}$,$\bar{k}$) over another. To this end, we restrict our attention to the best-fit values in Table~\ref{tab:summary2}.}

We next examine in detail the best-fits to the individual datasets. \textcolor{black}{The data for the change in the fraction of the original lattice extent for four different experiments are shown in Fig.~\ref{fig:fits}. It is important to note from Table~\ref{tab:summary} that the lattices for the four chosen experiments have very different spatial extents, which explains the differing time scales for contraction. In fact, it appears that the rate at which the fraction of the original extent of a lattice shrinks is inversely proportional to its original extent. This suggests that the contraction rate of the lattices are related to their initial sizes.}

We begin by considering the experiments performed by \citet{Bell1979} that yielded the first reported observation of FPCL contraction. In these experiments, different numbers of fibroblast cells were embedded in free-floating collagen lattices of varying ECM density, which contained fetal bovine serum. The fraction of the original length was then measured at various times over the course of several days. In the following, we consider their set of results that were obtained using $7.5\times 10^{6}$ fibroblast cells embedded in a lattice of initial diameter $53$mm. It was observed that this lattice subsequently contracted to around $10\%$ of its original diameter in around $9$ days. We find that the best fit of our numerical results to the experimental data can be obtained by using $\bar{\theta}=0.89$, $\bar{k}=106.62$ (see Fig.~\ref{fig:fits}a).

We next consider the experiments of \citet{Talas1997}, in which the difference between the effects of normal and recessive dystrophic epidermolysis bullosa fibroblasts were investigated. For the case of normal fibroblasts, $2.5\times 10^{5}$ cells were embedded in a lattice with a diameter of approximately $2.18$~cm. It was observed that the lattice contracted to around $37\%$ of its original area in around $7$ days. We find that the best fit of our numerical results to the experimental data can be obtained by using $\bar{\theta}=5.10$, $\bar{k}=1294.92$ (see Fig.~\ref{fig:fits}b).

Next, we consider the experiments of \citet{Feng2003}, in which the mechanical properties of contracted collagen gels were investigated. Here, $2.5\times 10^{6}$ fibroblast cells were embedded within a lattice of initial diameter $100$mm. It was observed that the lattice contracted to around $13\%$ of its original diameter by the end of the experiment, with most of the contraction occurring within the first few days. In this case, we find that the best fit of our numerical results to the experimental data can be obtained by using $\bar{\theta}=16.14$, $\bar{k}=183.85$ (see Fig.~\ref{fig:fits}c).

Finally, we consider the results of \citet{Guidry1985}, obtained when $10^{5}$ fibroblasts were placed on an attached lattice of diameter $12$mm. It was observed that this lattice contracted to about $24\%$ of its original thickness in around $1$ day. As the presence of protomyofibroblasts was not reported in these experiments, it is likely that the observed contraction is primarily due to the activity of fibroblasts and hence our model can be used to approximate this behaviour. We simulate the effect of adding cytochalasin D by allowing $N(X,\,T)$ to take the form (\ref{eq:N_inh}). Results obtained using $\bar{\theta}=0.33$, $\bar{k}=2.97$, which gave the best fit to the experimental data, are shown in Fig.~\ref{fig:fits}d. We find that in addition to the contraction of this gel, our model can capture the observed partial re-expansion.

\section{Discussion}
\label{S:discussion}

In this work, we develop a $1$-D morphoelastic model to describe the evolution of the diameter (width) of a free-floating (attached) collagen lattice that is contracted by fibroblasts. We fit numerical solutions of the model to previously obtained experimental data by optimizing two free parameters. In addition to being able to capture the contraction of the lattice, our model closely describes the partial re-expansion of the lattice observed in the experiment by \citet{Guidry1985}. Hence, this is to our knowledge the first mathematical model that explains how the permanent contraction of such lattices partially persists when reorganisation is inhibited by killing the fibroblasts or by otherwise preventing them from altering the lattice. We achieve this by explicitly taking into account the continual changes to the zero stress state of a material in response to a prescribed rate of growth and explicitly describes the evolution of this state. The partial re-expansion of the lattice is an example of a class of phenomena related to changes in the underlying tissue structure. As the theoretical framework developed here is flexible and versatile, it could potentially be used to model a range of other biological processes that involve internal remodelling of a tissue's mechanical structure, by making certain assumptions about the dependence of growth on other physical parameters.

Note that our framework for stress and strain is developed in Eulerian coordinates (in Appendix~\ref{S:Theory}). Although a Lagrangian framework can lead to equations with fixed boundaries and a useful variational structure (\textcolor{black}{see}, for example~\citet{RobertsCM} and \citet{GonzalezCM}) it is typically only appropriate in situations where the zero stress state does not evolve. However, as described in \citet{Yavari2010}, the introduction of a changing zero stress state leads to new terms that need to be carefully incorporated into the energy balance equation. Moreover, in cases where the zero stress state is close to the current state, but quite different from the initial state, Eulerian coordinates enable aspects of small deformation theory, such as the linear stress-strain relationship, to be applied.

On deriving the relations for stress and strain in Eulerian coordinates, we then revert to Lagrangian coordinates (in \ref{SS:extendedmodel}) in order to avoid the problems inherent in a system with a moving boundary. To do this, we exploit the fact that a conversion to Lagrangian coordinates corresponds to a conversion to characteristic variables (as diffusion is absent in our system), and we are thus able to simplify our model to a system of ordinary differential equations.

It should be noted that we have assumed in our model that the zero stress state is always uniform within each \textcolor{black}{unit} and that the expression for $g(x,\,t)$ also gives a uniform rate of change to the zero stress state.
\textcolor{black}{Against this,} it might be argued that the only region where the zero stress state is changing (due to the direct effect of fibroblast action) is the region where $-l_c < x < l_c$.
\textcolor{black}{However,} experiments have been performed where cells were cultured on top of lattices instead of throughout lattices, and these still showed relatively uniform contraction \citep{Grinnell1984,Guidry1985,Guidry1986}. \textcolor{black}{This indicates that while cell migration may be slow \citep{Grinnell1984,Ascione2016}, there is sufficient cell movement (or perhaps cells rearrange collagen over sufficiently large volumes) for it to be reasonable to assume that the change in the zero stress state is uniform in space.}

\textcolor{black}{Despite the fact that cell movement is quite limited over the time scale of FPCL contraction, a natural extension to our model would be to consider the case where cell density is nonuniform, and where there is active cell movement over the time-scale of contraction. Extensions like this would be necessary for developing models of wound healing based on the principles presented in this model. When cell motility is taken into consideration, the transformation to Lagrangian coordinates would no longer convert the partial differential equations of the original model to ordinary differential equations, and it is likely that it would be more straight forward to solve the original Eulerian equations.}


The morphoelastic approach detailed in this paper, where an evolution law is developed for the effective strain but a conventional stress-strain relationship is used,  could potentially be applied to several other biological processes including soft tissue growth, arterial remodelling, aneurysm development, morphogenesis, initiation of stretch marks and solid tumour growth. Furthermore, the permanent contraction observed in some pathological scars instead suggests that morphoelastic changes to the underlying extracellular matrix are particularly significant during the wound healing process. Hence, an important potential application of this theory is the development of a mechanochemical model of dermal wound healing. Indeed, it would be fruitful to build on recent theoretical models of the wound healing process that have considered aspects of the interplay between fibroblasts, keratinocytes and growth factors \citep{Menon2012}, the complementary r\^{o}les of TGF-$\beta$ and tissue tension \citep{Murphy2011a,Murphy2012}, the interplay between these components and ECM deformation~\citep{Valero2014}, and their r\^{o}le in hypertrophic scar formation~\citep{Koppenol2017a} and wound retraction~\citep{Koppenol2017b}.

In addition, there are several possible ways \textcolor{black}{in which} this model could be extended to \textcolor{black}{describe} other observed aspects of the behaviour of FPCLs. For instance, there have been experiments in which force measurement devices were attached to the collagen lattice \citep{Kolodney1992,Brown1996,Marenzana2006} in order to measure the total contractile force exerted by the cells. In each case, the force measurement device provides a finite spring-like resistance to the contraction of the lattice, and so the contraction process could be described using our model by replacing the stress-free boundary condition in (\ref{eq:stressboundary}) with one that relates the stress at $x = l(t)$ to the displacement at the contracting edge. However, as the increased elastic tension in such a lattice could lead fibroblasts to modulate into protomyofibroblasts \citep{Tomasek2002} that increase the stress, it may also be necessary to make the cell-associated stress, $\sigma_c$, a function of elastic stress. Indeed, this modulation can also occur in stress-relaxed FPCLs \citep{Tomasek1992}, which causes the lattice to contract rapidly when released. In the additional presence of TGF-$\beta$, protomyofibroblast cells in such lattices will further modulate into myofibroblasts \citep{Desmouliere1993,Tomasek2002,Gabbiani2003,Desmouliere2005}. In order to capture the effect of these highly contractile cells, our model could be modified via the addition of a new species for protomyofibroblasts (or myofibroblasts), a new chemical species, TGF-$\beta$, and including a conversion term in the equation for the fibroblasts. \textcolor{black}{Furthermore, as mentioned earlier, the effect of heterogeneity in gel compaction could also be investigated using the theoretical framework that we describe.}

However, it is important to note that although our morphoelastic model closely captures the important features of the permanent contraction of FPCLs, the approach used here can only truly be justified in $1$-D Cartesian co-ordinates (see Appendix~\ref{SS:MultiDecomp} for more details). Consequently, although our model can provide a good description of attached lattices, a complete description of cylindrical, free-floating lattices would require significant \textcolor{black}{modifications to the constitutive laws used for the mechanics of the lattice} (albeit with only minor adjustments to the reaction-diffusion equations). \textcolor{black}{This could still be achieved using the same principles described in Sec.~\ref{SS:cellcontract} if appropriate modifications are made; for example, we could consider a three-dimensional array of units, each containing a single cell, and each cell could be treated as a sphere of body stresses pulling inward. By homogenising such a system, we could develop a three-dimensional constitutive law for growth analogous to \eqref{growthrate}; however, any such rule would be significantly more complicated than the one-dimensional rule developed in this manuscript.}

An additional complication in this regard is that there remains considerable ambiguity surrounding the precise specification of $3$-D laws for the evolution of the zero stress state (for an overview on recent advances in this area, see~\citet{Kuhl2014}). As noted in \citet{Ambrosi2011}, there has been little success in using thermodynamic arguments to develop general frameworks for morphoelasticity. Furthermore, there are uniqueness issues surrounding the multiplicative decomposition of the deformation gradient and it is difficult to ensure that any phenomenological evolution law is appropriately observer-independent. Although some effort has gone into resolving these problems (especially in the engineering literature -- see, for example, \citet{LubardaElastoPlast} and \citet{Xiao2006}), the resulting models are often densely expressed and difficult to apply to biological morphoelasticity. As noted in \textcolor{black}{chapters $4$ and $5$ of} \citet{HallThesis}, some progress can be made by considering possible three-dimensional generalisations of (\ref{IntegralGrowth}), but there is a need for further work in this area. We also note that there is still interest in developing constitutive models that could provide better mathematical descriptions of remodelling in soft biological tissues (see, for example, \citet{Comellas2016}).

Despite these challenges, the morphoelastic framework presented \textcolor{black}{in this manuscript} has significant advantages over other approaches to describing biological remodelling. In particular, phenomena like the permanent contraction of a collagen lattice, which our model can describe, are inaccessible to classical Kelvin-Voigt models and are very different from the stress-induced contraction observed in Maxwell models.

\textcolor{black}{The central achievement of our work lies in developing a theoretical framework for the mechanics of tissue remodelling that explicitly accounts for the action of cells to change the fundamental structure of a collagen lattice. By defining strain in terms of the difference between the current state and the zero stress state and considering how this strain changes according to both physical deformation and the action of cells, we find that morphoelasticity can be easily incorporated into a model of FPCL contraction. Indeed, this approach allows us to use conventional constitutive laws to relate the stress to the strain.} The $1$-D morphoelastic framework described in this paper provides us with a simple and meaningful technique to the describe some of the complexities of biological remodelling, and it has the potential to be \textcolor{black}{extended and used} in a wide range of other areas.

\begin{acknowledgements}
This research was primarily supported by the Australian Research Council's Discovery Projects funding scheme (project number DP0878011). SNM is supported by the IMSc Complex Systems Project ($12^{\rm th}$ Plan).
CLH acknowledges the support of the Mathematics Applications Consortium for Science and Industry, funded by the Science Foundation Ireland grant investigator award 12/IA/1683.

\end{acknowledgements}

\appendix

\section{Effective strain and contraction in a $1$-D morphoelastic body}
\label{S:Theory}

\subsection{The multiplicative decomposition of the deformation gradient}
\label{SS:MultiDecomp}

The idea that an observed deformation gradient tensor $\mb{F}$ could be expressed in terms of elastic and plastic tensor fields through a multiplicative decomposition was first introduced by \citet{Bilby1957} and \citet{Kroner1958,Kroner1959}. This was further developed by \citet{Stojanovic1964,Stojanovic1970} in the context of thermoelasticity and \citet{Lee1969} for the description of metal plasticity at large deformations, while \citet{Rodriguez1994} and \citet{CookThesis} were the first to utilize this in the context of biomechanics. The pioneering work of \citet{Rodriguez1994} and \citet{CookThesis} was later extended and expanded by Hoger and coworkers (see especially \citet{Chen2000}), Goriely and coworkers \citep{BenAmar2005,Goriely2007,Goriely2008}, Ambrosi and coworkers \citep{Ambrosi2004,Ambrosi2007a,Ambrosi2007b} and by \citet{VandiverThesis}. This biological work has developed alongside applications to thermoelasticity and plasticity, and achievements in these areas have informed each other. This is exemplified by the cross-disciplinary work of \citet{LubardaElastoPlast,Lubarda2004} and Rajagopal and coworkers (see, for example, \citet{Rajagopal2004}). For a comprehensive review of the history and applications of the multiplicative decomposition of deformation gradient, see \citet{Lubarda2004}.

In a recent comprehensive review of current work on modelling growth and remodelling, \citet{Ambrosi2011} describe a number of applications of the multiplicative decomposition of the deformation gradient, ranging from the remodelling of heart muscle to morphogenesis. The multiplicative decomposition of the deformation gradient is now being used in models of biomechanical phenomena ranging from tissue growth \citep{BenAmar2005} to the operation of the heart \citep{Goktepe2010,Rausch2011}, although it is important to note that this approach will only be valid when the tissue behaves elastically on the timescale of remodelling \citep{WynJones2012}, and that some authors have general reservations about the use of the multiplicative decomposition on theoretical grounds \citep{Xiao2006}. Indeed, \citet{Ambrosi2011} note that there are problems and ambiguities to be resolved when developing appropriate laws to describe the evolution of the growth part of the deformation gradient in response to remodelling. In the context of the present work, it is important to note that many of these difficulties are avoided as we restrict our analysis to the one-dimensional ($1$-D) Cartesian case, although it is still necessary to ensure that the constitutive relation is appropriate for the type of remodelling under consideration.

An accessible introduction to the use of the multiplicative decomposition in biological applications can be found in \citet{Goriely2011}, which begins with an analysis of a $1$-D growing material that is relevant to the research presented here.
It is important to note that a $1$-D body can never be residually stressed: it is impossible to encounter the situation in which the zero stress state cannot be achieved without introducing cuts. Moreover, ensuring observer-independence of time derivatives is much simpler along a single dimension, as it excludes the possibility of a rotating observer. We now develop a simple mathematical framework for modelling the growth or contraction of a $1$-D Cartesian morphoelastic body.

\subsection{Strain evolution}
\label{SS:Effstrain}

The deformation gradient is given by the scalar function
\[
 F(X,\,t) = \pdiff{x}{X}\,,
\]
Following the $1$-D version of equation (1) in \citet{Goriely2011}, we express this as the product
\begin{equation}
F = \alpha \, \gamma \,,
\label{eq:FGor}
\end{equation}
where  the elastic stretch $\alpha$ is the local size ratio between the current state and the zero stress state, and the growth stretch $\gamma$ is the local size ratio between the zero stress state and the initial state (the growth stretch).

In \citet{Goriely2011}, the constitutive relation for $1$-D growth is assumed to depend on the rate of growth $g(x,\,t)$:
\begin{equation}
 \pdiff{\gamma}{t} = g(x,\,t)\,, \label{GorielyGrowth}
\end{equation}
While \eqref{GorielyGrowth} is useful at small deformations (\ie when $F \approx 1$), it leads to  inconsistencies if the current state is significantly different from the initial state, as in the case of FPCL contraction. In order to obtain an equivalent of \eqref{GorielyGrowth} for large deformations, we first note that $g(x,\,t)$ should satisfy
\begin{equation}
 \rdiff{}{t} \int_{X_A}^{X_B} \gamma(X,\,t) \, \mathrm{d}X = \int_{x(X_A,\,t)}^{x(X_B,\,t)} g(x,\,t) \, \mathrm{d}x. \label{IntegralGrowth}
\end{equation}
That is, $g(x,\,t)$ should be defined with reference to the current configuration, but it should measure the rate of change of the zero stress state of any collection of material particles. It follows from \eqref{IntegralGrowth} that $g(x,\,t)$ is related to the material derivative of $\gamma(X,\,t)$:
\begin{equation}
 \mdiff{\gamma}{t} = F \, g(x,\,t), \label{gammaevol}
\end{equation}
Note that this reduces to \eqref{GorielyGrowth} when $F \equiv 1$.

Now, we expect that the stress at any point in the body will be related to the difference between the zero stress state and the current state. A plausible constitutive law that relates the stress, $\sigma$, to the elastic stretch, $\alpha$ is
\begin{equation}
 \label{EulerianConstitutive}
 \sigma = E \, \left(1 - \alpha^{-1}\right)\,,
\end{equation}
where $E$ is the Young's modulus.
This is analogous to Hooke's law for a linear elastic material, but uses an Eulerian rather than a pseudo-Lagrangian measure of strain, since
\[
e^{E} \equiv 1 - \alpha^{-1} = \lim_{\Delta x \rightarrow 0} \frac{\Delta x - \Delta z}{\Delta x}, \text{ while }  e^{L} \equiv \alpha - 1 = \lim_{\Delta x \rightarrow 0} \frac{\Delta x - \Delta z}{\Delta z}\,,
\]
where $\Delta x$ and $\Delta z$ relate to the changes in the current and zero stress states, respectively. In cases where the current state is close to the zero stress state, and hence $\alpha \approx 1$, we see that $e^{E}\approx e^{L}$ and (\ref{EulerianConstitutive})
is equivalent to other plausible constitutive laws, such as those in \citet{Goriely2011}:
\begin{gather}
 \sigma = E \, (\alpha - 1), \label{LagrangianConstitutive}\\[6pt]
 \sigma = \frac{E}{3} \, \left(\alpha^2 - \alpha^{-1}\right), \label{NeoHookeanConstitutive}
\end{gather}
Experimental observations indicate that most of the change in size of a contracting FPCL is due to the permanent rearrangement of fibres by fibroblasts \citep{Guidry1985}. Hence, it is appropriate to use \eqref{EulerianConstitutive} and assume a linear relationship between stress and strain, rather than the nonlinear model \eqref{NeoHookeanConstitutive}. Moreover,
\eqref{EulerianConstitutive} has an interesting advantage over \eqref{LagrangianConstitutive} and \eqref{NeoHookeanConstitutive}, namely that the evolution of Eulerian strain in response to growth can neatly be expressed as an advection equation with a source term that is independent of $e^E$.

In order to see this, we substitute (\ref{eq:FGor}) into \eqref{gammaevol} to obtain
\begin{equation}
 F \, \mdiff{}{t} \, \alpha^{-1} + \alpha^{-1} \, \mdiff{F}{t} = F \, g(x,\,t)\,. \label{AlphaEvolTemp}
\end{equation}
Since $F^{-1} \, D F/D t = \partial v/\partial x$
where $v$ is the velocity and where $\partial v/\partial x$ is the velocity gradient, it follows that $\alpha^{-1}$ satisfies the equation
\begin{equation*}
 \pdiff{}{t} \alpha^{-1} + \pdiff{}{x} \left(v \, \alpha^{-1}\right) = g(x,\,t)\,, \label{AlphaInvEvol}
\end{equation*}
and using $e^{E} \equiv 1 - \alpha^{-1}$ we thus obtain the mechanical model for a morphoelastic solid with small effective strain
\begin{equation}
\label{StrainEvolutiona}
\pdiff{e^E}{t} + \pdiff{}{x} \left(e^E \, v \right)  = \pdiff{v}{x} - g(x,\,t)\,,
\end{equation}
Note that thus far we have made the assumption that the relation between stress and strain is purely elastic
\begin{equation*}
\label{StrainEvolutionb}
\sigma = E \, e^E\,.
\end{equation*}
However, as discussed in Sec.~\ref{SS:cellcontract}, this can easily be extended to viscoelastic bodies through the use of a Kelvin-Voigt viscoelastic constitutive law. In Sec.~\ref{S:application} of the main text, we use this formulation together with (\ref{StrainEvolutiona}) to derive a set of governing equations for the contraction of a $1$-D morphoelastic body.


\section{Derivation of the spatial and temporal transformations between coordinate systems}
\label{sec:AppTransf}

In one spatial dimension, we use $X$ to represent the Lagrangian coordinate and $x$ to represent the Eulerian coordinate. At any given time $t$, there will be a one-to-one mapping from the initial configuration to the current configuration. Thus, we can always write $X = X(x,\,t)$ and $x = x(X,\,t)$; moreover, the fact that particles are not permitted to move through each other implies that ${\partial x}/{\partial X} > 0$.

Now, the Eulerian displacement gradient is the spatial derivative of $u(x,\,t) =x-X(x,\,t)$, i.e.
\begin{equation}
w{(x,t)}=1-\pdiff{X}{x}\,.
\label{eq:A0}
\end{equation}
Similarly, the Lagrangian displacement gradient is
\begin{equation*}
W(X,\,T)=\pdiff{}{X}\left(x-X\right)=\frac{1}{1-w}-1\,.
\end{equation*}
We thus have the relation
\begin{equation}
\label{eq:wW}
1+W=\frac{1}{1-w}\,.
\end{equation}
Using the chain rule, the Eulerian spatial derivative is
\begin{equation*}
\pdiff{}{x}\equiv\pdiff{X}{x}\pdiff{}{X}+\pdiff{T}{x}\pdiff{}{T}\,.
\end{equation*}
The derivative $\partial T/\partial x$ is equal to zero, and using (\ref{eq:A0}) we obtain the expression
\begin{equation}
X=x-\int_{0}^{x}\,w(\xi,t) \mathrm{d}\xi\,.
\label{eq:A1}
\end{equation}
Furthermore, using (\ref{eq:A0}) and (\ref{eq:wW}), we obtain the following transformation for the spatial derivative
\begin{equation}
\pdiff{}{x}\equiv \frac{1}{1+W}\pdiff{}{X}\,.
\label{eq:A2}
\end{equation}
We similarly use the chain rule to obtain the following expression for the Eulerian temporal derivative
\begin{equation*}
\pdiff{}{t}\equiv\pdiff{X}{t}\pdiff{}{X}+\pdiff{T}{t}\pdiff{}{T}\,.
\end{equation*}
The derivative $\partial T/\partial t$ is equal to one, and so using (\ref{eq:A1}) we have
\begin{equation*}
\pdiff{}{t}\equiv\pdiff{}{t}\left(\int^{x}_{0} 1 - w(\xi,t)\,\mathrm{d}\xi\right)\pdiff{}{X}+\pdiff{}{T}\,.
\end{equation*}
Using (\ref{eq:weqn}), we have
\begin{equation*}
\pdiff{}{t} \equiv \int^{x}_{0} \pdiff{}{\xi}\left(w(\xi,t)\,v(\xi,t)-v(\xi,t)\right) \,\mathrm{d}\xi\pdiff{}{X}+\pdiff{}{T}\,,
\end{equation*}
which, on using (\ref{eq:wW}) and the fact that $v(0,\,t)=0$, yields
\begin{equation*}
\pdiff{}{t}\equiv\pdiff{}{T}-\frac{v}{1+W}\pdiff{}{X}\,.
\end{equation*}
Now, from (\ref{eq:velocity}), we have $v\,(1-w)=\partial{u}/\partial{t}$. Since $u\equiv U$, using the above expression in conjunction with (\ref{eq:wW}), this yields
\begin{equation*}
v = (1+W)\,\left( V-\frac{v}{1+W}\,W \right)\,,
\end{equation*}
where we have used the definitions of $V$ and $W$. This implies that $v\equiv V$, and so we obtain the following transformation for the temporal derivative
\begin{equation}
\pdiff{}{t}\equiv\pdiff{}{T}-\frac{V}{1+W}\pdiff{}{X}\,.
\label{eq:A3}
\end{equation}
Now, from (\ref{eq:weqn}) we have $\partial{w}/\partial{t} + \partial{(w\,v)}/\partial{x} = \partial{v}/\partial{x}$. Using (\ref{eq:wW}), (\ref{eq:A2}) and (\ref{eq:A3}), this reduces to the following relation between the Lagrangian displacement gradient and velocity:
\begin{equation}
\label{eq:WV}
\pdiff{W}{T}=\pdiff{V}{X}\,.
\end{equation}
We use this expression in Sec.~\ref{SS:extendedmodel} of the main text to derive our morphoelastic model of FPCL contraction.


\bibliographystyle{spbasic}      

\end{document}